# GRANDES FRAUDES Y GOBIERNOS CORPORATIVOS EN LA ECONOMÍA DESDE MEDIADOS DEL SIGLO XX

*INÉS MARTÍN DE SANTOS*[1]

> Les escribo a ustedes como profesional de una disciplina desgraciada. La teoría económica, tal como se enseña desde los años 80, ha fracasado miserablemente en punto a entender las fuerzas determinantes de la crisis financiera. Conceptos centrales, como los de las "anticipaciones racionales" y la "disciplina de mercado", así como la "hipótesis de la eficiencia de los mercados", llevaron a los economistas a sostener que la especulación estabilizaría los precios, que los vendedores actuarían para proteger su propia reputación, que podía confiarse en el *caveat emptor* y que, por consecuencia, no podría darse un fraude generalizado. No todos los economistas creyeron eso. Pero la mayoría, sí. (James K. Galbraith. *Por qué los economistas no vieron el fraude financiero*. The Huffington Post, 12 de octubre 2009, p. 1. Testimonio ante el Senado de los EEUU 06/06/10).

## 1. PREÁMBULO

El término *fraude* (fraud: wrongful or criminal deception intended to result in financial or personal gain) se puede presentar, al menos, desde dos perspectivas diferentes: una parte ética o deontológica como comportamiento deshonesto, y otra parte jurídica como delito contra la legislación vigente de un gobierno. Estos dos aspectos son diferentes, pero hay que tenerlos en cuenta como aspectos complementarios o, dicho de otro modo, no debieran ser analizados de manera excluyente.

Elaborar una tipología de fraudes ocuparía mucho tiempo y probablemente nunca obtendría un consenso general. Someramente, en la vida social, podemos considerar demasiadas clases de fraudes, pero los más vinculantes para nosotros, en esta ocasión, son solapadamente los fraudes científicos y sobre todo los fraudes de índole económica, en los que se centra este estudio.

El fenómeno no es nuevo sin embargo, a veces resulta comprometedor e inapropiado considerar fraudes algunas creencias populares como sucede con las de los bestiarios, antecedentes de la actual Zoología, o las cuestiones de carácter metodológico como la opinión los enemigos del álgebra cartesiana en el siglo XVII, convencidos de que los problemas matemáticos podían seguir resolviéndose mediante tradicionales procedimientos geométricos.

---

[1]. Profesora Ayudante Doctora. Departamento de Administración Financiera y Contabilidad. Universidad Complutense de Madrid. E-mail inesmartin@ucm.es

Otra cosa es faltar a la verdad conscientemente o presumir de haber recurrido a las fuentes originales sin haberlo hecho. Ya en la España del siglo XVIII, José Cadalso (1772) en *Los eruditos a la violeta* aludía a la arrogancia de ciertos sabios indocumentados: "Cualquier libro que os citen, decid que ya lo habéis leído, y examinado" (p. 68).

En esta línea, uno de los mayores escándalos mundiales fue el imaginario hombre de Piltdown (1912) que suponía el eslabón perdido de la cadena de los homínidos y tuvo engañada a la ciencia durante 40 años.

Más recientemente, uno de los fraudes más llamativos fue el del físico Jan Hendrik Schön, doctor por la Universidad de Constanza y trabajador de los laboratorios Bell de New Jersey. El hecho se descubrió porque Schön y sus coautores publicaban un artículo por semana y dio que pensar cuando se advirtió que en sus artículos se repetían los mismos gráficos en diferentes experimentos. La comisión investigadora que revisó los trabajos en el año 2005 comprobó 16 amaños en 24 artículos; lo más sorprendente fue, además, que 8 de ellos aparecieron en la revista *Science*.

El plagio y el autoplagio son, asimismo, tipos de fraudes corrientes que ni siquiera han podido frenar a veces los derechos de la propiedad intelectual y de la propiedad industrial.

Alertadas por las informaciones falsas, muchas revistas científicas de prestigio están revisando de manera retrospectiva sus publicaciones y se están detectando aproximadamente unos 500 artículos con informaciones fraudulentas cada año.

## 2. INTRODUCCIÓN

Los fraudes tanto a la hacienda pública como a los particulares, de variados tipos y bajo diversas denominaciones, han aparecido en todo tiempo y lugar. Durante el pasado siglo, el más conocido fue la crisis norteamericana producida por el hundimiento de la bolsa de Nueva York, iniciada en 1929, que afectó no sólo a la economía norteamericana sino a gran parte del resto del mundo (la Gran Depresión) y que duró en Estados Unidos hasta la terminación de la Segunda Guerra Mundial (aproximadamente 1954, año de recuperación de la bolsa neoyorquina). Con el objetivo de prevenir futuras debacles a nivel macroeconómico, a partir de los años 50 se ha venido desarrollando la idea de Responsabilidad Social Corporativa (RSC) que en la práctica se materializa en los Gobiernos Corporativos (GC) de las grandes y, en menor medida, medianas empresas.

Los fraudes más destacados se han extendido por muchos otros países pero los más señalados se han producido en los grandes centros financieros (Peachey, 2011). España, como la mayor parte de los países, no ha sido ni es ajena a las fluctuaciones internacionales derivadas de las malas prácticas financieras.

Habida cuenta de experiencias previas, se considera y evalúa si los Gobiernos Corporativos se muestran suficientemente operativos para prevenir futuros desastres económicos a gran escala.

## 3.  LOS FRAUDES ECONÓMICOS

La teoría de los ciclos económicos es una de las constantes más comúnmente aceptadas en Economía y a veces asumidas como mal menor. Unas veces los cambios de ciclo se pueden producir de manera inopinada por inercia o agotamiento del modelo económico, como hasta hace poco se pensaba; pero otras veces son provocados por acontecimientos especiales que inciden en el desarrollo ordinario de la actividad económica a modo, por ejemplo, de un músico que desafinara en una orquesta: la interpretación de la pieza, por consiguiente, resultaría nefasta. Estos últimos fenómenos son los que más tienen que ver con los grandes fraudes que producen desastres principalmente de carácter macroeconómico.

El origen de los fraudes financieros es tan viejo como la historia de la humanidad ya que nace desde el mismo momento en que aparecen los préstamos de bienes y servicios, y pueden observarse vestigios de estos hechos tanto en la India como en China o la antigua Mesopotamia entre los siglos XX y XVIII antes de nuestra era.

La financiación ha sido la base para el sostenimiento tanto de los imperios como de los países soberanos, principalmente para afrontar los gastos derivados de los conflictos bélicos. El problema fundamental del préstamo es la devolución de la deuda. Entre las grandes quiebras medievales podemos destacar, por ejemplo, la de la familia Peruzzi. En la Edad Moderna, tanto en España (Caro Baroja, 1985) como en el resto de Europa (Delumeau, 1983) continuaron los desfalcos.

En cuanto a los fraudes económicos, es preciso recordar que la primera gran bancarrota de la edad moderna se produjo en España en 1557 a comienzos del reinado de Felipe II a causa del exagerado endeudamiento de su padre Carlos I para afrontar los gastos del imperio. A partir de entonces, la deuda exterior ha sido constantemente deficitaria llegando a casos extremos como el de las desamortizaciones decimonónicas para poder evitar la suspensión de pagos.

España, con catorce casos, encabeza el número de veces que un país se ha declarado en suspensión de pagos. Le siguen, en cuanto al número de veces y relevancia: Venezuela y Ecuador (11), Brasil (10), Francia y México (9), Alemania, Argentina y El Salvador (8), Colombia, Uruguay y Portugal (7), Estados Unidos y Rusia (6).

La declaración de quiebra técnica puede parecer una medida fácil de adoptar, pero sólo se debe aplicar en casos excepcionales e irremediables ya que el gran problema aparecerá con posterioridad cuando el país abandone una necesaria y empobrecedora situación de autarquía y desee volver a competir en los mercados internacionales.

## 4.  LOS DESFALCOS INDIVIDUALES

Interesa conocer algunos casos concretos de estafas para cotejarlos con las penas que les fueron impuestas. Retrotrayendo nuestra mirada un poco antes del período seleccionado en el este estudio, una de las personas más perjudiciales fue Carlo Ponzi, quien en 1919 ideó un sistema peculiar de estafa prometiendo unas rentabilidades del 50% y hasta del 100% a corto plazo, un modelo similar a la estructura piramidal y a la estructura multinivel. Por su primera estafa sólo llegó a cumplir tres años de cárcel. Fue reincidente.

Bernard Madoff cometió otro fraude similar al de Ponzi sobre una aparente capitalización manipulada a base de ingeniería económica. Lo sorprendente fue que el engaño duró unos 16 años entre 1992 y 2008. Con 71 años, fue condenado a 150 años de cárcel. Sentencia ilógica.

Frank Abagnale, en los años 90, fue precoz en el arte de la falsificación de talones bancarios combinada con los cambios de identidad personal. Se le conmutó la condena por su colaboración con la policía. Lo preocupante fue su declaración de que hoy tendría 4000 veces más posibilidades de estafar y esta afirmación no debe extrañar dado que se prevé que las redes que funcionen con tecnología 5G así como los repositorios big data, independientemente de las indudables mejoras para la comunicación, puedan propiciar una *ciberdelincuencia*.

Nicholas Leeson, ejecutivo de Banca Baring hundió esta entidad debido a sus desacertadas gestiones en el mercado de futuros de Singapur. Se declaró culpable de engaño a sus superiores y a la bolsa de este país. La condena fue de seis años y medio que no llegó a cumplir íntegramente.

Richard Fuld y Josep Gregory, últimos directivos de Lehman Brother's declararon en quiebra este banco de inversión y gestión de activos el 15 de setiembre del 2008. La responsabilidad de sus otros responsables se ha ido diluyendo con el tiempo y finalmente se produjo una situación de impunidad. Los perjudicados de esta bancarrota fueron tanto los accionistas como los acreedores. Además, el caso Lehman Brother's puso en tela de juicio la credibilidad de las agencias de calificación o empresas auditoras.

En relación con España, Nieves Carrera (2011) ha estudiado los fraudes más importantes de las últimas décadas a partir de 1980: Caja Rural de Jaén, Banesto, Gescartera, Promotora Social de Viviendas e Iniciativas de Gestión de Servicios (PSV), Afinsa y Fórum Filatélico (este último creado por el ex-fraile agustino Jesús Fernández Prada). Habría que añadir Sofico (1974. En este caso, escapa al período seleccionado por la investigadora), Rumasa y Acciones Preferentes de Caja Madrid (hoy Bankia).

Los pequeños fraudes no parecen menos relevantes, sobre todo si se tiene en cuenta que son señas de identidad de una determinada sociedad. Se pueden consultar en la base de datos INEbase. He seleccionado aquellos que tienen que ver con la actividad económica, a saber:

| | 2018 | 2017 | 2016 | 2015 | 2014 | 2013 |
|---|---|---|---|---|---|---|
| 13 Contra el patrimonio y el orden socioeconómico | 142.426 | 136.992 | 122.647 | 74.790 | 64.620 | 60.645 |
| 13.6 Defraudaciones | 23.252 | 21.953 | 17.736 | 10.202 | 9.166 | 7.876 |
| 13.13 Delitos societarios | 81 | 96 | 77 | 106 | 92 | 67 |
| 13.14 Receptación y blanqueo de capitales | 3.161 | 3.021 | 3.000 | 2.733 | 2.495 | 2.009 |
| 14 Contra la Hacienda Pública y Seguridad Social | 1.396 | 1.002 | 937 | 938 | 712 | 647 |
| 18 Falsedades | 8.499 | 7.783 | 7.921 | 7.453 | 7.376 | 6.770 |
| 18.1 Falsificación de moneda y efectos timbrados | 157 | 178 | 127 | 146 | 154 | 150 |
| 19 Contra la Administración Pública | 1.527 | 1.678 | 1.692 | 1.591 | 1.548 | 1.588 |
| 19.8 Fraudes y exacciones ilegales | 21 | 41 | 39 | 25 | 26 | 13 |

Fuente: INEbase. Tabla. Última actualización hasta 2018.

Como se puede observar en el gráfico siguiente, la evolución de los delitos va en aumento, y aquí solamente figuran aquellos fraudes juzgados y condenados. Más inquietante que las cifras y los porcentajes es la evolución de la curva en sentido ascendente.

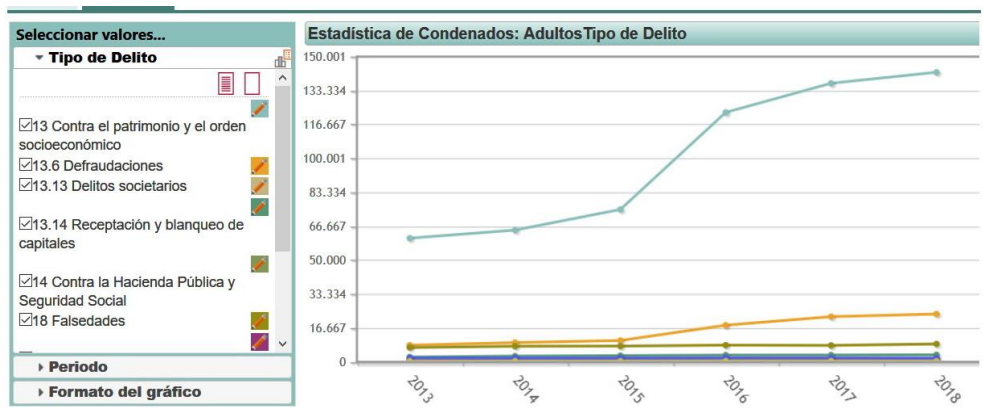

Fuente: INEbase, Gráfico. Última actualización hasta 2018.

## 5. LAS CRISIS DE MAYOR ENVERGADURA

Como ya he adelantado anteriormente, la aparición de un gran fraude puede ocasionar una crisis. La crisis más conocida del mundo contemporáneo fue la producida en Estados Unidos conocida como el Crack del 29 o la Gran Depresión, propiciada fundamentalmente por una sobreproducción manufacturera, una concesión crediticia sin garantías y, en definitiva, una sobrevaloración de activos en bolsa poco acorde con la economía real. A pesar de haber tenido un origen nacional, tiene su incidencia en la mayor parte del mundo debido al comercio internacional. La segunda guerra mundial atemperó esta crisis y fue en buena parte la solución que hizo salir de ella a Estados Unidos aproximadamente hacia 1954.

En 1973 se produjo la primera crisis del petróleo. En este caso no se trataba de una crisis financiera diversificada sino de una materia prima de primer orden. Fue, además, una crisis provocada de manera arbitraria por parte de los países árabes productores de petróleo al suprimir la exportación de esta materia a los países defensores de Israel en la guerra de Yom Kipur contra Siria y Egipto. La decisión árabe fue un duro golpe sobre todo para Estados Unidos y sus aliados europeos al propiciar un tremendo coste de productos derivados del petróleo y la exagerada inflación.

En 1990 nuevamente el aumento de los precios del petróleo, juntamente con la burbuja inmobiliaria japonesa, sacudieron la estabilidad del sistema financiero internacional.

A partir de entonces hasta la actualidad, también ha habido crisis importantes, aunque de menor alcance mundial: la mejicana de 1994, la asiática de 1997, la rusa de 1998 y la argentina de 2001.

Entre 1997 y 2001 cabe destacar la crisis de las empresas conocidas genéricamente como las *puntocom*; algunas de estas empresas realizaron enormes inversiones en el sector informático y de las telecomunicaciones. Adquirieron en bolsa unas cotizaciones si no ficticias, al menos bastante alejadas de su valor real. En 2001 esta burbuja reventó y en 2002 aún aumentó su devaluación en el índice Nasdaq produciendo enormes pérdidas a sus inversores.

Además de la ya citada caída de Lehman, el mayor escándalo financiero fue el del gigante norteamericano Enron, empresa dedicada al sector energético que no solo mostraba unos activos abultados sino que también realizó sobornos y cometió delitos fiscales. La desacertada calificación de Arthur Andersen hizo desaparecer prácticamente casi del todo esta empresa auditora.

La quiebra Enron está considerada en las publicaciones de carácter divulgativo como la más grande de la historia tras la Gran Depresión aunque si nos atenemos a las cifras que presenta Jaramillo (2013, s.p.), fue mucho mayor la catástrofe de WorlCom, casi el doble. Independientemente de su perjuicio a los accionistas, podemos también asegurar que su contagio fue notorio en la economía internacional y sirvió de modelo para una escalada de quiebras similares dentro y fuera de Estados Unidos, asentadas en la falsificación de sus estados de cuentas, sueldos multimillonarios para los directivos y, en definitiva, lo que se ha dado en llamar la aplicación práctica de la ingeniería contable y financiera.

En España el caso más significativo, entre las *puntocom*, fue el de Terra la compañía que acabó quebrando en 2005 tras seis años de vicisitudes.

La crisis financiera más reciente, y antesala de la actual, provocada por la pandemia internacional del virus covid-19, se venía gestando desde 2006 y estalló en 2008, precedida de las hipotecas *subprime* o hipotecas basura en EEUU como el anteriormente citado caso Lehman Brother's que contagió a la banca de muchos otros países y afectó principalmente al sector inmobiliario. Dio lugar a una abundante producción cinematográfica y literaria entre cuyas obras destaca el bestseller *Griftopia* (2010).

En definitiva, si hacemos un recuento de las crisis más importantes acaecidas en el planeta durante el último medio siglo, constatamos la cifra de diez (incluyendo la actual). Es, por consiguiente, un dato preocupante para el sostenimiento del sistema capitalista, que es el predominante en casi todo el mundo, sobre todo a partir del deterioro de la hegemonía de Estados Unidos en el siglo XXI (Dabat, 2009) y de las directrices estadounidenses que puedan afectar a otros países.

## 6. LOS PERÍODOS CRÍTICOS

La mayor parte de las crisis se produce con el advenimiento de las revoluciones o cambios tecnológicos que en general se muestran más eficientes por la absorción o menor necesidad de mano de obra y el consiguiente paro. El caso de las *puntocom* que he nombrado sería un ejemplo de esta clase de crisis inevitables.

Sin embargo, observamos que junto a estos fenómenos hay que añadir el factor que voy a denominar *voluntad humana,* encaminada sobre todo al desarrollo de las operaciones financieras.

Referirse a las finanzas es hablar de economía dinámica, es decir de inversión pero también de deuda, y un endeudamiento desproporcionado suele ser el germen de cualquier crisis, y remontar una crisis puede suponer una duración de varias décadas. Este tipo de crisis es evitable.

El barómetro de medir las crisis principalmente es la bolsa. Antes he mencionado la crisis del 29, la más larga de todas por el momento, pero el 19 de octubre de 1987 también se resintieron los cimientos de gran parte de las bolsas del mundo (la española perdió nada menos que un 31% de su valor ese día). El motivo de esta debacle no fueron las medidas fiscales de Estados Unidos tendentes a mermar la caída de beneficios, como demostró Shiller (2003), sino un presentimiento general de los inversores a la caída, motivado por el uso del apalancamiento en las compras y el excesivo endeudamiento generalizado.

Desde una perspectiva diacrónica, los fraudes y las crisis aparecidas a partir de la última década del siglo pasado han sido más frecuentes. Las tecnologías basadas en la Informática posibilitan y facilitan las tareas laborales pero también el crecimiento de las estafas, y los gobiernos deben arbitrar medidas adecuadas para impedir o al menos frenar este desajuste del sistema. Una solución apropiada, entre otras, es la instauración de gobiernos corporativos al menos en las grandes empresas y entidades financieras.

## 7. LOS GOBIERNOS CORPORATIVOS

En la actualidad, estamos leyendo con frecuencia frases como economía circular, desarrollo sostenible, economía colaborativa (por ejemplo, la popular plataforma Airbnb), producción ecológica, responsabilidad social corporativa, gobierno corporativo y equivalentes, es decir observamos que se extiende cada vez más una voluntad regeneracionista para el eficiente aprovechamiento de los recursos naturales. Los accionistas de las mayores entidades gestoras de fondos de inversión (con BlacRok a la cabeza) están mostrando, asimismo, un interés especial por conectar con esta tendencia a la hora de tomar decisiones.

*Grosso modo*, se puede decir que el Gobierno Corporativo (Corporate Governance) es la materialización de la Responsabilidad Social Corporativa o, dicho de otra forma, la inclusión real en los órganos rectores de la empresa de los denominados *stakeholders* entendiendo como tales no sólo presidente y cuerpo directivo sino también instituciones sin ánimo de lucro, entidades gubernamentales, accionistas, representantes de los propios asalariados e incluso de los consumidores, es decir miembros propios o ajenos a la empresa con capacidad influyente en la planificación de la actividad empresarial.

El establecimiento de gobiernos corporativos tiene como en tantas otras ocasiones sus defensores y detractores. De hecho, en el título de algunas obras (Martín Fernández, 2019; Turullols, 2019) ya es indicativo que se haya incluido la palabra *buen* puesto que da a entender que puede haber malos o inadecuados gobiernos corporativos como también la adaptación a la Responsabilidad Social Corporativa puede anunciarse como mera táctica de mercadotecnia para promocionar las ventas.

La figura de los gobiernos corporativos está pensada sobre todo para empresas con una estructura compleja y una facturación extraordinaria, de todas formas también puede incluir grandes empresas de carácter familiar (Rayza Mirelle, F.N. et al., 2019).

La puesta en marcha de un gobierno corporativo en las empresas no es tan novedosa como puede parecer y viene a relevar y hacer efectiva la idea de la Responsabilidad Social Corporativa (RSC) desarrollada inicialmente por Bowen (1953) y que cuenta en nuestro país con un nutrido grupo de seguidores, organización de congresos monográficos y publicación de manuales en constante renovación, entre los que me atrevo a destacar el de García del Junco (2018).

Algunos proyectos caracterizados por el espíritu de Responsabilidad Social Corporativa, además de dar muestras de solidaridad social, han presentado unos resultados satisfactorios como es el caso de los microcréditos concedidos Muhammad Yunus en Bangladesh, sin embargo no siempre suponen ventajas (Vargas Escudero, 2006), incluso Yunus intentó repetir el mismo experimento en Perú y fracasó. Las razones de este fracaso probablemente haya que encontrarlas en la idiosincrasia de los pueblos, un factor al que hay que dar más importancia de la que a menudo se da.

La instauración de gobiernos corporativos en las empresas es una de las herramientas posiblemente más poderosas para hacer frente a los grandes desfalcos que puedan producirse en el mundo de las finanzas. Pero esta figura no cuenta con el respaldo de muchos empresarios y además puede entrañar ciertas dificultades para un ágil funcionamiento de las decisiones empresariales. Por ejemplo, establecer un gobierno corporativo en el mercado de derivados apenas es operativo (Turegeldiyev, 2014).

Los gobiernos corporativos, en todo caso, parecen insuficientes para atajar los problemas de fraude, puesto que por regla general sus funciones comienzan y terminan en la adopción de decisiones y no suelen realizar seguimientos de gestión. En la actualidad, se están desarrollando en el ámbito informático algoritmos que supuestamente son más potentes que el mero control manual de los expertos. Un interesante estudio reciente ha expuesto las modalidades más empleadas llevadas a cabo (Ramírez-Alpízar, 2020).

## 8. DIMENSIÓN SOCIAL DE LOS FRAUDES

La incidencia de un fraude económico en la vida social depende de la magnitud del mismo. La víctima siempre es la hacienda pública y en último caso la sociedad, pero el perjuicio afectará en mayor o menor grado a las personas relacionadas con el origen de la estafa.

La ventaja que hoy día tiene la sociedad es una mayor información sobre los fraudes recibida tanto a través de los medios oficiales o públicos como privados (*mass media*) y desde luego las populares redes sociales, principales forjadoras de la opinión pública.

En el sistema económico actual los demandantes de productos o servicios, por el hecho de estar más informados que en otro período histórico anterior, desempeñan un papel fundamental en el mercado como fuerza activa. Las decisiones de los consumidores están mejor forjadas que en épocas pasadas y ello se debe en gran parte al desarrollo de la Responsabilidad Social Corporativa, antesala de los Gobiernos Corporativos.

La creciente preocupación social por los fraudes económicos ha generado una inmensa literatura. La extensión de grandes fraudes tanto en el ámbito de comercio tradicional como en el del e-commerce ha sido y es objeto de multitud de publicaciones, algunas con un acentuado carácter divulgativo (Taibbit, 2010); otras con excelentes compilaciones académicas (Jones, 2011), otras derivadas de directivas gubernamentales (v.g. Comisión Europea, 2001), e incluso la repercusión del tema ha llegado también a documentales de gran éxito (Ferguson, 2010, (Sington, 2011), además de las creaciones cinematográficas: *Wall Street* (2010), *Margin Call* (2011), *Too big to fail* (2011), *The Big Short* (2015),…

Cada día que pasa es mayor el número de personas que, por ejemplo, a la hora de hacer la compra no sólo se fija en el precio de los alimentos, también mira la procedencia, si son productos transgénicos o sin tratar, si se ha empleado mano de obra barata o ilegal, si la empresa reinvierte parte de sus ganancias en mejorar el medio ambiente, en fomentar la conciliación familiar.

## 9. CONCLUSIONES

Si se acepta la consideración del dinero como deuda, se puede afirmar que el sistema financiero internacional en la actualidad aún no está preparado para afrontar una previsible crisis motivada principalmente por la dicotomía entre economía real y economía financiera o virtual. El escepticismo cunde incluso cuando se trata de inversiones en economía sostenible.

A pesar del fenómeno de la globalización, el mundo actual se encuentra a gran distancia de resolver los problemas derivados de los grandes fraudes para el desarrollo del crecimiento económico sostenible.

La concentración del capital en pocas manos es uno de los mayores riesgos para la posible reiteración de las crisis económicas. Se apunta como necesaria la anti-globalización como una posibilidad para mantener el espíritu de la libre competencia.

Las benévolas sentencias de los tribunales a algunos de los defraudadores no contribuyen a alejar el fantasma del fraude como tampoco a la desaparición de los paraísos fiscales. En la actualidad, seguimos lejos de cumplir los deseos del Presidente de la Segunda República Española Manuel Azaña de que los defraudadores devuelvan las ganancias conseguidas ilegítimamente.

El comercio electrónico si bien es capaz de mostrar su eficiencia aminorando costes, tampoco es menos cierto que puede propiciar fraudes a pequeña y gran escala habida cuenta de que permite mayor nivel de enmascaramiento de los defraudadores.

Desde la perspectiva diacrónica, se observa que las crisis económicas son cada vez más frecuentes e incidentes siempre en ámbito financiero; lo que obliga a replantear un modelo económico a escala internacional en el que se produzca un mayor peso de la política económica de los gobiernos sobre el poder de las compañías multinacionales en el contexto de la globalización. Esta política debe dar prioridad al sostenimiento de los recursos naturales. Una de las formas posibles es el desarrollo de gobiernos corporativos en las empresas de mayor incidencia en el PIB de los países.

En el contexto de la Responsabilidad Social Corporativa, los Gobiernos Corporativos se apuntan como uno de los resortes fundamentales para frenar los grandes fraudes empresariales pero su eficiencia parece insuficiente a falta de una normativa internacional y el desconocimiento de fuerzas ocultas en lo que se viene conociendo como ingeniería fiscal y financiera.

La aplicación de unas políticas liberales de manera poco ortodoxa, que no se corresponden en muchos casos con la doctrina de los maestros, está ocasionando enormes desajustes en el sistema, está provocando verdaderas brechas sociales en la distribución de la renta y está minando el actual sistema capitalista.

La popularización de la información sobre las crisis financieras a través no sólo de la literatura sino sobre todo de documentales y películas de cine y televisión está formando una conciencia social cada vez más crítica con los escándalos financieros de gran influencia que aboca a las grandes empresas a la necesidad de implementar gobiernos corporativos como una de las fórmulas imprescindibles para el sostenimiento del sistema económico internacional.

Se insta al desarrollo internacional de leyes de transparencia parecidas a la Ley española 19/2013, de 9 de diciembre, de transparencia, acceso a la información pública y buen gobierno.

## 10. BIBLIOGRAFÍA